%% file: article.tex
\documentclass[12pt]{article}
\usepackage{epsfig}
\usepackage{graphicx}
\usepackage{pstricks}
\usepackage{epsf}
\usepackage{epsfig}
\usepackage{amsmath,amsfonts,amsthm, amssymb,latexsym}

\newcommand{\beq}{\begin{equation}}
\newcommand{\eeq}{\end{equation}}
\newcommand{\bcn}{\begin{center}}
\newcommand{\ecn}{\end{center}}

\newcommand{\lsim}{\lower0.5ex\hbox{$\; \buildrel < \over \sim \;$}}

\input econfmacros-csqcd3.tex
\def\Title#1{\begin{center} {\Large {\bf #1} } \end{center}}

\begin{document}

\Title{Thermal Evolution of Rotating Neutron Stars}

\bigskip\bigskip


\begin{raggedright}

{\it Rodrigo Negreiros\index{}\\
Instituto de F\'isica\\
Universidade Federal Fluminense\\
Av. Gal. Milton Tavares S/N Gragoat\'a\\
Niter\'oi, RJ\\
Brazil\\
{\tt Email: negreiros@if.uff.br}}
\bigskip

{\it Stefan Schramm\index{}\\
Frankfurt Institute for Advanced Studies\\
Goethe University, Ruth Moufang Str. 1\\
60438 Frankfurt,\\
Germany\\
{\tt Email:schramm@th.physik.uni-frankfurt.de}}
\bigskip

{\it Fridolin Weber\index{}\\
Department of Physics\\
San Diego State University\\
5500 Campanile Drive,\\
San Diego, California 92182,\\
USA\\
{\tt Email:fweber@sdsu.edu}}
\bigskip\bigskip

\end{raggedright}

\section{Introduction}

The cooling of neutron stars have been used by many
authors \cite{Schaab1996,Page2004,Page2006,Page2009,Blaschke2000,Grigorian2005,
  Blaschke2006,Page2011,Yakovlev2011} as a way of probing the
internal composition of these objects. Such studies rely on the fact that the
physical quantities relevant for the cooling (specific heat, thermal
conductivity, and neutrino emissions) strongly depends on the microscopic
composition, and thus different models lead to different thermal evolution. The
predicted thermal evolution is compared with observed data, with the ultimate
goal of constraining the microscopic model \cite{Page2004,Page2006}. Recent
studies along those lines \cite{Page2011,Yakovlev2011} has linked the observed
thermal behavior of the compact object in CasA \cite{Heinke2010} to the possible
onset of hadronic superfluidity in the core of neutron stars. A possible
alternative explanation to the cooling behavior of CasA has been proposed by the
authors in \cite{Negreiros2011}, where the observed data was explained as the
late onset of the Direct Urca process, triggered by the compression that
accompanies a spinning down neutron star.

In the usual approach, studies of the thermal evolution of neutron stars are
performed assuming a spherically symmetric object. As pointed out in
\cite{Negreiros2011}, however, rotation may play an important role on the
thermal evolution of neutron stars. Whereas in \cite{Negreiros2011} the
effects of rotation in the microscopic composition (and its consequences to
cooling) where considered, in the work presented here we complement that study
\cite{Negreiros2011}, and investigate the macroscopic aspects of the cooling of
rotating neutron stars.

In this work we consider the thermal evolution of rigidly rotating neutron
stars. In order to perform such study we first calculate the structure of
rotating objects, which is considerably more complicated than that of spherical
objects. The structure of rotating neutron stars is obtained by
solving Einstein's equation for a rotationally deformed fluid distributions
\cite{Weber,Glendenning2000}. The numerical method used is based on the the KEH
method \cite{Komatsu1989,Cook1992,Stergioulas1995}. The equation of state
used for computing the neutron star structure and composition is a simple
relativistic mean field (RMF) model, with parameter set
G300 \cite{Glendenning1989}. With the structure of rotating neutron stars
computed, we calculate the thermal evolution of these objects. In order to do
so, we re-derive the thermal evolution equations to account for the metric of a
rotating object. The cooling of neutron stars with different frequencies
is then calculated. We show that the cooling of the star strongly depends on
the frequency of the object, with higher frequencies stars showing a substantial
temperature difference between the equator and poles.

\section{2D thermal evolution}
We review the here the thermal evolution equations for a rotating neutron star,
as discussed in \cite{Negreiros2012}. The metric of a rotationally deformed
fluid can be written as \cite{Weber},
\begin{eqnarray}
ds^2 = - e^{2 \nu} dt^2 + e^{2 \phi} (d\varphi - N^\varphi dt)^2 + e^{2
\omega} (dr^2 + r^2 d\theta^2) ,
\label{eq:3.1}
\end{eqnarray}
where $e^{2\phi} \equiv e^{2(\alpha + \beta)} r^2 \sin^2\theta$ and $e^{2
  \omega} \equiv e^{2(\alpha-\beta)}$.  The quantities $\nu$, $\phi$ and
$\omega$ denote metric functions, and $N^\varphi$ accounts for frame dragging
caused by the rotating fluid. All these functions are to be computed
self-consistently from Einstein's field equation, $G^{\bar\alpha \bar\beta} =
8 \pi T^{\bar\alpha \bar\beta}$, where $T^{\bar\alpha \bar\beta}$ denotes the
fluid's energy momentum tensor.

For a uniformly rotating compact star ($\Omega=$const), the equations of
energy balance and transport can be reduce to
\begin{eqnarray}
  \partial_t \tilde T = - \, {e^{2 \nu \over{\Gamma^2}} } {\epsilon \over{
      C_V}} + 
 {1 \over{r^2 \sin\theta}} {e^{3\nu-\gamma-2\xi} \over \Gamma} 
{1 \over {C_V}}\times \quad \quad \quad \quad \quad \quad \nonumber & \\
\Bigl( \partial_r \left(
r^2 \kappa \sin \theta \, e^\gamma
\left( \partial_r \tilde T  
\right) \right) +
 {1 \over{r^2}} \partial_\theta \left( r^2 \kappa \sin \theta \,
e^\gamma
\left( \partial_\theta \tilde T  \right) \right) \Bigr) \, , & \\
\nonumber \label{eq:3.49}
\end{eqnarray}
with the definitions $r \sin\theta e^{-\nu+\gamma} = e^\phi$ and
$e^{-\nu+\xi} = e^{\alpha-\beta}$. In the above equation  $T$ is the
temperature, $\tilde T \equiv e^\nu T / \Gamma$, $\kappa$ is the thermal
conductivity, $C_V$ is the specific heat, $\epsilon$ is the neutrino emissivity,
and the Lorentz factor $\Gamma \equiv (1- U^2)^{-1/2}$ where $U$ is the four
velocity. 
The standard cooling equations of spherically symmetric, non-rotating neutron
stars are obtained from Eq.\ (\ref{eq:3.49}) for $\Omega=
0$ and $\partial_\theta \tilde T=0$ \cite{Weber}. In this work we solve Eq.\
(\ref{eq:3.49}) for the temperature distribution
$T(r,\theta;t)$ of non-spherical rotating neutron stars.  
The boundary condition are given by defining the heat flux at $r=0, R$, and
 at $\theta=0, \pi/2$, with $R$ denoting the stellar radius. The star's initial
temperature, $T(r,\theta;t=0)$, is chosen as $\tilde T \equiv
10^{11}$~K. 
 Eq. (\ref{eq:3.49}) is the equation that
needs to be solved numerically, where the macroscopic input (metric functions
and radius) is provided by solution of the structure of the rotating neutron
stars, and the microscopic input (specific heat, thermal conductivity and
neutrino emissivity) is derived from the equation of state. 

In this work we consider the following neutrino processes for the stellar core:
the direct Urca, modified Urca and bremsstrahlung processes.  A detailed review
of the emissivities of such processes can be found in
reference \cite{Yakovlev2001a}. In addition to the core we also consider
the standard processes that take place at crust of a neutron
star \cite{Yakovlev2001a}.



\section{Surface Temperature Evolution} 
We now present the results for the thermal evolution of rotating neutron stars.
In order to evaluate the impact of the 2D structure, we studied the thermal
evolution of three different stars, with the same central density, and different
frequencies. The properties of these stars are listed in table
\ref{table:stars}, where $M$ denotes the gravitational mass, $R_e$ the
equatorial radius, $R_p$ the polar radius, $r = R_p/R_e$ is the ratio between
polar and equatorial radii, and $\Omega$ the rotation frequency.
\begin{table}[b]
\begin{center}
\begin{tabular}{ccccc}\hline
$M/M_\odot$   & $R_e$ (km) & $R_p$ (km) & $r$ & $\Omega$ (hz)  \\
\hline
      & &  & \\
  1.28 & 13.25  & 13.11 & 0.99 & 148     \\ 
  1.34 & 13.85  & 12.49 & 0.90 & 488     \\ 
  1.48 & 15.21  & 11.50 & 0.75 & 755     \\ 
\end{tabular}
\caption{ Properties of the different stars whose thermal
evolution was investigated. All stars have a central density of 350 MeV/fm$^3$,
and were computed using a Relativistic Mean Field model with parameter set G300
\cite{Glendenning1989}}
\label{table:stars}
\end{center}
\end{table}

\begin{figure}[htb]
\begin{center}
\epsfig{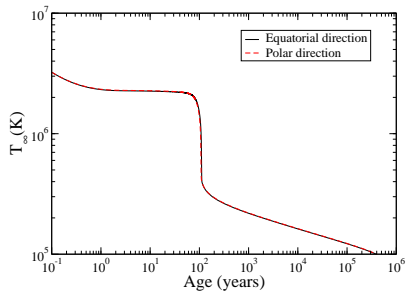}
\caption{Redshifted temperature as a function of
age of the pole and equator of the 148 hz star in table \ref{table:stars}.}
\label{Cool_1}
\end{center}
\end{figure}

Equation \ref{eq:3.49} was solved numerically for each star of table
\ref{table:stars}, and the results are shown in
figs.~\ref{Cool_1}--\ref{Cool_3}. In these figures we plot the redshifted
temperature of the poles and of the equator of the star. 

\begin{table}[b]
\begin{center}
\begin{tabular}{cc}
\epsfig{file=cooling_488.eps,height=1.5in}
  & 
\epsfig{file=cooling_755.eps,height=1.5in}
\end{tabular}
\caption{ Same as Fig.~\ref{Cool_1} but for the 488 hz (right) and 755 hz
(left) star.}
\label{Cool_3}
\end{center}
\end{table}

The cooling curves shown in Fig.~\ref{Cool_1} indicate that there is no
difference between the polar and equatorial temperatures. As indicated in table
\ref{table:stars}, for the frequency of this star (148 hz) the ratio between
polar and equatorial radii $\sim 0.99$. The matching of the polar and equatorial
temperatures indicates that the cooling of this object is the same as that of
a spherical object. As the frequency of the star increases (or equivalently
as $r$ decreases) one can see more significant differences between the
polar and equatorial temperatures, with the pole being slightly warmer than the
equator. Another notable difference can be seen at the sharp temperature drop at
$\sim 100$ years. Such temperature drop is associated with the thermal
coupling between the core and the crust of the star, and its magnitude depends
on whether or not fast neutrino emission processes are taking place.
The microscopic model used in this work allows the Direct Urca process for stars
with masses above 1.0 M$_\odot$, which explains the the sharp temperature drop
observed. Since this "knee" is associated with the thermal coupling between core
and crust \cite{Gnedin2001}, it is only natural that it happens at different
times for the polar and equatorial direction. This behavior is due to the
deformation of the star, which causes it to be flattened at the poles (leading
to a thinner crust in the polar direction), and elongated at the equator
(leading to a thicker crust). The thinner crust, combined with the low fluid
velocity in the polar region, allows the cold front (that originates in the
core) to reach this region before the equator, as observed in Figs.
\ref{Cool_3}. For lower frequencies (Fig. \ref{Cool_1}) the
structure approaches that of a spherically symmetric stars and thus this effect
vanishes.

\section{Conclusions}

It was the objective of this work to investigate in details the thermal
evolution of rotating neutron stars. Such investigation is
extremely important if one wants to understand the thermal evolution of systems
in which the spherical symmetry is broken like in highly magnetized neutron
stars, in spinning-down neutron stars, and/or of accreting objects. In order
to achieve our objective we studied the relatively simple case of a rotating
neutron stars with rigid rotation, and constant frequency. The metric of
rotating neutron stars is significantly different than that of spherically
symmetric objects, hence the general relativistic equations that govern the
thermal evolution were re-derived to account for the new metric. 

We have studied the thermal evolution of neutron stars with a purely
hadronic composition (the whole baryon octet) and that allows for the presence
of the direct Urca process. The microscopic model used leads to the so-called
"enhanced" cooling. This choice of equation of state was intentional, since it
allow us to study the energy transport inside the rotating neutron star
more clearly. After obtaining a clear understanding of the thermal evolution
of rotating compact stars, our study can be applied to system with
more sophisticated features like hadronic superfluidity, and quark matter for
instance. 

Our study shows that the 2D thermal evolution of neutron stars can in fact be
significantly different than that of spherically symmetric objects. We have
found that the deformation in a neutron star (in the case studied here due to
rotation) leads to non-uniform surface gravity, which in turns leads to
non-uniform surface temperatures. We have also found that the deformation of the
crust plays a major role for the thermal evolution of these objects. Now that we
have a better understanding of the heat transport inside a deformed neutron
star, we intend to extend this research to account for the effects described in
\cite{Negreiros2011}, where  spin-down compression, and its consequences to the
cooling of a neutron stars were discussed.  We also plan to include magnetic
field effects, since those also contribute to braking the spherical symmetry of
the object. It is also our intention to explore more sophisticated microscopic
models, that account for hadronic superfluidity, and possibly quark matter for
instance.

\end{document}

%% file: econfmacros-csqcd3.tex
\def\beq{\begin{equation}}
\def\eeq#1{\label{#1}\end{equation}}
\def\eeqn{\end{equation}}

\def\beqa{\begin{eqnarray}}
\def\eeqa#1{\label{#1}\end{eqnarray}}
\def\eeqan{\end{eqnarray}}

\let\bar=\overbar

\def\Dslash{\not{\hbox{\kern-4pt $D$}}}
\def\dslash{\not{\hbox{\kern-2pt $\del$}}}

\def\msb{{\bar{\ssstyle M \kern -1pt S}}}

\usepackage{fancyhdr,graphicx}

%% file: article.bbl
\begin{thebibliography}{99}



\expandafter\ifx\csname natexlab\endcsname\relax\def\natexlab#1{#1}\fi
\expandafter\ifx\csname bibnamefont\endcsname\relax
  \def\bibnamefont#1{#1}\fi
\expandafter\ifx\csname bibfnamefont\endcsname\relax
  \def\bibfnamefont#1{#1}\fi
\expandafter\ifx\csname citenamefont\endcsname\relax
  \def\citenamefont#1{#1}\fi
\expandafter\ifx\csname url\endcsname\relax
  \def\url#1{\texttt{#1}}\fi
\expandafter\ifx\csname urlprefix\endcsname\relax\def\urlprefix{URL
}\fi
\providecommand{\bibinfo}[2]{#2}
\providecommand{\eprint}[2][]{\url{#2}}

%

\bibitem{Schaab1996}
\bibinfo{author}{\bibfnamefont{C.}~\bibnamefont{Schaab}},
  \bibinfo{author}{\bibfnamefont{F.}~\bibnamefont{Weber}},
  \bibinfo{author}{\bibfnamefont{M.~K.}~\bibnamefont{Weigel}},
\bibnamefont{and}
  \bibinfo{author}{\bibfnamefont{N.~K.} \bibnamefont{Glendenning}},
  \bibinfo{journal}{Nuclear Phys A,} \textbf{\bibinfo{volume}{605}},
  \bibinfo{pages}{531,} (\bibinfo{year}{1996}).

\bibitem{Page2004}
\bibinfo{author}{\bibfnamefont{D.}~\bibnamefont{Page}},
  \bibinfo{author}{\bibfnamefont{J.}~\bibnamefont{Lattimer}},
  \bibinfo{author}{\bibfnamefont{M.}~\bibnamefont{Prakash}},
\bibnamefont{and}
  \bibinfo{author}{\bibfnamefont{A.~W.} \bibnamefont{Steiner}},
  \bibinfo{journal}{The Astrophysical Journal Supplement Series,}
  \textbf{\bibinfo{volume}{155,}}, \bibinfo{pages}{623}
(\bibinfo{year}{2004}).

\bibitem{Page2006}
\bibinfo{author}{\bibfnamefont{D.}~\bibnamefont{Page}},
  \bibinfo{author}{\bibfnamefont{U.}~\bibnamefont{Geppert}},
\bibnamefont{and}
  \bibinfo{author}{\bibfnamefont{F.}~\bibnamefont{Weber}},
  \bibinfo{journal}{Nuclear Physics A,} \textbf{\bibinfo{volume}{777}},
  \bibinfo{pages}{497} (\bibinfo{year}{2006}).

\bibitem{Page2009}
\bibinfo{author}{\bibfnamefont{D.}~\bibnamefont{Page}},
  \bibinfo{author}{\bibfnamefont{J.}~\bibnamefont{Lattimer}},
  \bibinfo{author}{\bibfnamefont{M.}~\bibnamefont{Prakash}},
\bibnamefont{and}
  \bibinfo{author}{\bibfnamefont{A.~W.} \bibnamefont{Steiner}},
  \bibinfo{journal}{The Astrophysical Journal,}
\textbf{\bibinfo{volume}{707}},
  \bibinfo{pages}{1131} (\bibinfo{year}{2009}).

\bibitem{Blaschke2000}
\bibinfo{author}{\bibfnamefont{D.}~\bibnamefont{Blaschke}},
  \bibinfo{author}{\bibfnamefont{T.}~\bibnamefont{Klahn}},
\bibnamefont{and}
  \bibinfo{author}{\bibfnamefont{D.~N.}~\bibnamefont{Voskresensky}},
  \bibinfo{journal}{The Astrophysical Journal,}
\textbf{\bibinfo{volume}{533}},
  \bibinfo{pages}{406} (\bibinfo{year}{2000}).

\bibitem{Grigorian2005}
\bibinfo{author}{\bibfnamefont{H.}~\bibnamefont{Grigorian}},
  \bibinfo{author}{\bibfnamefont{D.}~\bibnamefont{Blaschke}},
\bibnamefont{and}
  \bibinfo{author}{\bibfnamefont{D.}~\bibnamefont{Voskresensky}},
  \bibinfo{journal}{Physical Review C,} \textbf{\bibinfo{volume}{71}},
  \bibinfo{pages}{045801} (\bibinfo{year}{2005}).

\bibitem{Blaschke2006}
\bibinfo{author}{\bibfnamefont{D.}~\bibnamefont{Blaschke}},
  \bibinfo{author}{\bibfnamefont{D.}~\bibnamefont{Voskresensky}},
  \bibnamefont{and}
  \bibinfo{author}{\bibfnamefont{H.}~\bibnamefont{Grigorian}},
  \bibinfo{journal}{Nuclear Physics A,} \textbf{\bibinfo{volume}{774}},
  \bibinfo{pages}{815} (\bibinfo{year}{2006}).

\bibitem{Page2011}
\bibinfo{author}{\bibfnamefont{D.}~\bibnamefont{Page}},
  \bibinfo{author}{\bibfnamefont{M.}~\bibnamefont{Prakash}},
  \bibinfo{author}{\bibfnamefont{J.}~\bibnamefont{Lattimer}},
\bibnamefont{and}
  \bibinfo{author}{\bibfnamefont{A.~W.}~\bibnamefont{Steiner}},
  \bibinfo{journal}{Physical Review Letters,}
\textbf{\bibinfo{volume}{106}},
  \bibinfo{pages}{081101} (\bibinfo{year}{2011}{\natexlab{a}}).

\bibitem{Yakovlev2011}
\bibinfo{author}{\bibfnamefont{D.~G.} \bibnamefont{Yakovlev}},
  \bibinfo{author}{\bibfnamefont{W.~C.~G.} \bibnamefont{Ho}},
  \bibinfo{author}{\bibfnamefont{P.~S.} \bibnamefont{Shternin}},
  \bibinfo{author}{\bibfnamefont{C.~O.} \bibnamefont{Heinke}},
  \bibnamefont{and} \bibinfo{author}{\bibfnamefont{A.~Y.}
  \bibnamefont{Potekhin}}, \bibinfo{journal}{Monthly Notices of the
Royal
  Astronomical Society,} \textbf{\bibinfo{volume}{411}},
\bibinfo{pages}{1977}
  (\bibinfo{year}{2011}).

\bibitem{Heinke2010}
\bibinfo{author}{\bibfnamefont{C.~O.} \bibnamefont{Heinke}}
\bibnamefont{and}
  \bibinfo{author}{\bibfnamefont{W.~C.~G.} \bibnamefont{Ho}},
  \bibinfo{journal}{The Astrophysical Journal,}
\textbf{\bibinfo{volume}{719}},
  \bibinfo{pages}{L167} (\bibinfo{year}{2010}).

\bibitem{Negreiros2011}
R. Negreiros, S. Schramm, and F. Weber, Phys. Lett. B, {\bf 718}, 1176 (2013).


\bibitem{Weber}
\bibinfo{author}{\bibfnamefont{F.}~\bibnamefont{Weber}},
  {\it {Pulsars as astrophysical laboratories for nuclear and
  particle physics}} (\bibinfo{publisher}{Institute of Physics},
  \bibinfo{address}{Bristol}, \bibinfo{year}{1999}),
\bibinfo{edition}{1st} ed.

\bibitem{Glendenning2000}
\bibinfo{author}{\bibfnamefont{N.~K.} \bibnamefont{Glendenning}},
  {\bibinfo{title}{ \it {Compact stars: nuclear physics, particle
physics, and
  general relativity}}} (\bibinfo{publisher}{Springer},
\bibinfo{year}{2000}),
  \bibinfo{edition}{1st} ed.


\bibitem{Komatsu1989}
\bibinfo{author}{\bibfnamefont{H.}~\bibnamefont{Komatsu}},
  \bibinfo{author}{\bibfnamefont{Y.}~\bibnamefont{Eriguchi}},
\bibnamefont{and}
  \bibinfo{author}{\bibfnamefont{I.}~\bibnamefont{Hachisu}},
  \bibinfo{journal}{Royal Astronomical Society, Monthly Notices, }
\textbf{\bibinfo{volume}{237}},
\bibinfo{pages}{355} (\bibinfo{year}{1989}).

\bibitem{Cook1992}
\bibinfo{author}{\bibfnamefont{G.~B.} \bibnamefont{Cook}},
  \bibinfo{author}{\bibfnamefont{S.~L.} \bibnamefont{Shapiro}},
  \bibnamefont{and} \bibinfo{author}{\bibfnamefont{S.~A.}
  \bibnamefont{Teukolsky}}, \bibinfo{journal}{The Astrophysical
Journal,}
  \textbf{\bibinfo{volume}{398}}, \bibinfo{pages}{203}
(\bibinfo{year}{1992}).

\bibitem{Stergioulas1995}
\bibinfo{author}{\bibfnamefont{N.}~\bibnamefont{Stergioulas}}
\bibnamefont{and}
  \bibinfo{author}{\bibfnamefont{J.~L.} \bibnamefont{Friedman}},
  \bibinfo{journal}{The Astrophysical Journal,}
\textbf{\bibinfo{volume}{444}},
  \bibinfo{pages}{306} (\bibinfo{year}{1995}).


\bibitem{Glendenning1989}
\bibinfo{author}{\bibfnamefont{N.~K.} \bibnamefont{Glendenning}},
  \bibinfo{journal}{Nuclear Physics A,} \textbf{\bibinfo{volume}{493}},
  \bibinfo{pages}{521} (\bibinfo{year}{1989}).

  
\bibitem{Negreiros2012}
R. Negreiros,\ S. Schramm,\ and F. Weber, Physical Review D, {\bf 85}, 104019
(2012)

\bibitem{Yakovlev2001a}
\bibinfo{author}{\bibfnamefont{D.~G.} \bibnamefont{Yakovlev}},
  \bibinfo{author}{\bibfnamefont{A.~D.} \bibnamefont{Kaminker}},
  \bibinfo{author}{\bibfnamefont{O.~Y.} \bibnamefont{Gnedin}},
  \bibnamefont{and}
\bibinfo{author}{\bibfnamefont{P.}~\bibnamefont{Haensel}},
  \bibinfo{journal}{Physics Reports,} \textbf{\bibinfo{volume}{354}},
  \bibinfo{pages}{1} (\bibinfo{year}{2001}).

\bibitem{Gnedin2001}
\bibinfo{author}{\bibfnamefont{O.~Y.} \bibnamefont{Gnedin}},
  \bibinfo{author}{\bibfnamefont{D.~G.} \bibnamefont{Yakovlev}},
  \bibnamefont{and} \bibinfo{author}{\bibfnamefont{A.~Y.}
  \bibnamefont{Potekhin}}, \bibinfo{journal}{Monthly Notices of the
Royal
  Astronomical Society,} \textbf{\bibinfo{volume}{324}},
\bibinfo{pages}{725}
  (\bibinfo{year}{2001}).
  


  
%
%
%
%
%
%
%
%
%
%
%
%
%
%
%
%
%
%
%
%
%
%
%
%
%
%
%
%
%
%
%
%
%
%


\end{thebibliography}
